\documentclass[superscriptaddress,preprintnumbers,twocolumn,prd,nofootinbib]{revtex4-1}

\usepackage{hyperref}
\usepackage{amsmath}
\usepackage{graphicx}
\usepackage{xspace}

\newcommand{\JPsi}{$\mathrm{J}\hspace{-.08em}/\hspace{-.14em}\psi$\xspace} 
\newcommand{\pt}{$p_T$\xspace}
\newcommand{\pT}{p_T}
\newcommand{\tev}{TeV}
\newcommand{\mev}{MeV}
\newcommand{\nb}{nb}
\newcommand{\invfb}{fb$^{-1}$}

\newcommand{\eq}[1]{Eq.~\eqref{eq:#1}}

\newcommand{\pythia}{{\sc pythia}\xspace}
\newcommand{\herwig}{{\sc herwig}\xspace}
\newcommand{\powheg}{{\sc powheg}\xspace}
\newcommand{\madgraph}{{\sc madgraph}\xspace}

\begin{document}

\title{Progress in Double Parton Scattering Studies}

\author{Sunil Bansal}
\affiliation{Department of Physics, University of Antwerp, 2020 Antwerp,
Belgium}

\author{Paolo Bartalini*\,}
\affiliation{Department of Physics, Central China Normal University, 430079 Wuhan, China}

\author{Boris Blok}
\affiliation{Department of Physics, Technion - Israel Institute of Technology, Haifa, Israel}

\author{Diego Ciangottini}
\affiliation{Dipartimento di Fisica, Universit\`a degli Studi di Perugia, 06100 Perugia, Italy}
\affiliation{INFN, sezione di Perugia, via A. Pascoli, 06100 Perugia, Italy}

\author{Markus Diehl}
\affiliation{Theory Group, Deutsches Elektronen-Synchrotron (DESY), 22607 Hamburg, Germany}

\author{Fiorella M.~Fionda}
\affiliation{Dipartimento interateneo di Fisica, Universit\`a degli Studi di Bari, Via Amendola 173, 70126 Bari, Italy}
\affiliation{INFN, sezione di Bari, Via E. Orabona n. 4, 70125 Bari, Italy}

\author{Jonathan R.~Gaunt}
\affiliation{Theory Group, Deutsches Elektronen-Synchrotron (DESY), 22607 Hamburg, Germany}

\author{Paolo Gunnellini}
\affiliation{Dipartimento di Fisica, Universit\`a degli Studi di Perugia, 06100 Perugia, Italy}
\affiliation{INFN, sezione di Perugia, via A. Pascoli, 06100 Perugia, Italy}

\author{Tristan Du Pree}
\affiliation{Universit\'e Catholique de Louvain, 1348 Louvain-la-Neuve, Belgium}

\author{Tomas Kasemets}
\affiliation{Nikhef, Theory Group, Science Park 105, 1098 XG, Amsterdam, The Netherlands}
\affiliation{Department of Physics and Astronomy, VU University, De Boelelaan 1081, 1081 HV, Amsterdam, the Netherlands}

\author{Daniel Ostermeier}
\affiliation{Institut f\"ur Theoretische Physik, Universit\"at Regensburg, 93040 Regensburg, Germany}

\author{Sergio Scopetta}
\affiliation{Dipartimento di Fisica, Universit\`a degli Studi di Perugia, 06100 Perugia, Italy}
\affiliation{INFN, sezione di Perugia, via A. Pascoli, 06100 Perugia, Italy}

\author{Andrzej Si\'odmok}
\affiliation{School of Physics and Astronomy, The University of Manchester, Manchester, M13 9PL, U.K.}

\author{Alexander M.~Snigirev}
\affiliation{Skobeltsyn Institute of Nuclear Physics, Moscow State University, 119991 Moscow, Russia}

\author{Antoni Szczurek}
\affiliation{Institute of Nuclear Physics PAN, PL-31-342 Cracow, Poland}
\affiliation{University of Rzesz\'ow, PL-35-959 Rzesz\'ow, Poland}

\author{Daniele Treleani*}
\affiliation{Dipartimento di Fisica dellÕUniversitˆ di Trieste, Trieste, Italy}
\affiliation{INFN, Sezione di Trieste, Strada Costiera 11, Miramare-Grignano, I-34151 Trieste, Italy}

\author{Wouter J.~Waalewijn*\,}
\affiliation{Nikhef, Theory Group, Science Park 105, 1098 XG, Amsterdam, The Netherlands}
\affiliation{ITFA, University of Amsterdam, Science Park 904, 1018 XE, Amsterdam, The Netherlands}

\date{\today}

\begin{abstract}
An overview of theoretical and experimental progress in double parton scattering (DPS) is presented. The theoretical topics cover factorization in DPS, models for double parton distributions and DPS in charm production and nuclear collisions. On the experimental side, CMS results for dijet and double \JPsi production, in light of DPS, as well as first results for the 4-jet channel are presented. ALICE reports on a study of open charm and \JPsi multiplicity dependence.
\end{abstract}

\maketitle

%
%

\section{Progress in the Theory of \\ Double Parton Scattering}

\subsection{Introduction}

Theoretical predictions for double parton scattering (DPS) require a factorization theorem for the cross section, in order to separate the two short-distance collisions from the long-range physics of the incoming protons. The partonic cross section of the hard scatterings is perturbatively calculable. The momenta of the quarks and gluons inside the proton are described by non-perturbative (double) parton distribution functions (PDFs), which must be modeled or extracted from data. 

If the two hard scatterings are independent and the two incoming partons in each proton are completely independent, the DPS cross section simplifies to 
\begin{equation} \label{eq:pocket_formula}
\sigma_\text{DPS} = \frac{\sigma_1 \sigma_2}{S \sigma_\text{eff}}
\,,\end{equation}
with $\sigma_1$ and $\sigma_2$ the standard cross sections of the individual scatterings and $S$ a symmetry factor. This leaves a single nonperturbative parameter $\sigma_\text{eff}$, which only affects the total DPS rate. For certain applications this approximation is sufficient, but one would also like to know the limitations and proper generalization of \eq{pocket_formula}. This is reflected in the variety of topics discussed within the DPS track at this workshop:
\begin{itemize}
\item Progress in factorization for DPS
\item Double parton correlations in proton models
\item DPS in charm cross sections
\item DPS in nuclear collisions
\end{itemize}

\subsection{Progress in Factorization}

A factorization analysis of DPS~\cite{Diehl:2011tt,Diehl:2011yj,Manohar:2012jr} reveals a large number of effects that are not included in the ``pocket formula" in \eq{pocket_formula}:
\begin{itemize}
\item Correlations between the two momentum fractions, the transverse separation of partons and/or flavor
\item Spin correlations between the partons
\item Color correlations between the partons
\item Interferences in fermion number
\item Interferences in flavor
\end{itemize}
Given the large number of effects and the suppressed nature of double parton scattering, it is important to identify which of these will likely dominate in experiments and how they will affect final-state distributions. Imposing that double PDFs satisfy momentum and quark-number sum rules, captures some of the correlations between momentum fractions and flavor~\cite{Gaunt:2009re}. (The validity of these sum rules is related to the 1v1 contribution discussed at the end of this section and thus remains unclear.) Spin correlations can be studied through the angular correlations they induce in the final state~\cite{Kasemets:2012pr}. Another handle is provided by the renormalization group evolution (RGE) of the double PDFs, since they need to be evolved to the scale of the hard scattering. For example, this implies that color correlations and interferences in fermion number are Sudakov suppressed and not important at sufficiently high scales~\cite{Mekhfi:1988kj,Diehl:2011yj,Manohar:2012jr}. 

A common assumption in deriving factorization formulae is that the contribution of gluons in the so-called Glauber region cancels out, which has only been rigorously proven for a few single parton scattering processes. This is investigated in Ref.~\cite{Diehl:2014tba} for double Drell-Yan with incoming scalars, fully differential in the momenta of the produced $Z/\gamma^*$. Specifically, the full NLO calculation is compared to the corresponding result obtained from the factorization formula (without Glauber contributions), which agree for the two and three point contributions. The four point contribution is still being studied.

The effect of the RGE of the double PDF on spin correlations and correlations between $x_i$ and $\mathbf{b}$ was investigated in Ref.~\cite{Diehl:2014vaa}. Correlations between $x_i$ and $\mathbf{b}$ reduce as the energy of the hard scattering increases, but only slowly. Spin correlations are washed out faster, particularly at small $x_i$ and for linearly polarized gluons. Given the typically low energy scales of hard scatterings in DPS these correlations are likely important. This study did not include contributions from single PDFs mixing into double PDFs, which would reintroduce correlations.

Aspects of this so-called 2v1 contribution to the cross section, involving one double PDF and one single PDF plus perturbative splitting, were studied in Ref.~\cite{Gaunt:2012dd}. Due to a different dependence on the transverse separation $\mathbf{b}$, it was found that $\sigma_\text{eff}$ should be larger for this contribution than for the term with two double PDFs. In the 2v1 contribution there can be cross talk (interference) between the two partons in the double PDF, which is not Sudakov suppressed. However, it has a small color factor and rather limited space for evolution, since it is cut off at the scale of the perturbative splitting rather than the hard scattering.

This role of this 2v1 contribution was also studied in Ref.~\cite{Blok:2013bpa}, which presented numerical results for integrated DPS cross sections for jet production at Tevatron and LHC energies. Here an uncorrelated approximation similar to \eq{pocket_formula} is used, but the correlation between the momentum fraction and impact parameter of the individual partons is kept. The initial state is then described by generalized double parton distributions, which are the product of two single generalized parton distributions, as obtained from HERA. It is found that the that the perturbative correlations induced by the 2v1 interactions significantly contribute to resolving the longstanding puzzle of an excess of multi-jet production events in the back-to-back kinematics, observed at the Tevatron and LHC. 

In the 1v1 contribution to DPS, only one parton is taken out of each proton and these partons both undergo a perturbative splitting before their daughter particles collide. This contribution is susceptible to double counting with loop corrections to single parton scattering~\cite{Cacciari:2009dp,Diehl:2011yj,Gaunt:2011xd,Gaunt:2012dd,Ryskin:2011kk,Blok:2011bu}. It has been proposed to leave out this contribution from the DPS cross section~\cite{Manohar:2012pe,Blok:2013bpa}, but that leads to a joint double PDF for both protons, breaking factorization in the usual sense of the word. This issue is important to resolve to obtain a consistent framework and because of the large correlation between partons produced by a perturbative splitting.

Looking forward, it is important to identify which of the many possible correlations are mostly likely to be probed in experimental studies  in order to move beyond testing whether $\sigma_\text{eff}$ is a universal quantity. Phrased differently, it will be crucial to identify regions of phase space or processes which (likely) have a different value of $\sigma_\text{eff}$, and to infer properties of the correlation structure of the proton from the corresponding measurements.

\subsection{Correlations in Proton Models}

A study of the double PDF in the bag model~\cite{Chang:2012nw} indicates that correlations between the two momentum fractions $x_i$, as well as spin correlations, are fairly large in the valence region $x_i \gtrsim 0.1$ at low scales. Correlations between $x_i$ and the transverse separation $\mathbf{b}$ were found to be small.

An analysis within constituent quark models, which contains two particle correlations without the need of an additional prescription (unlike Ref.~\cite{Chang:2012nw}), was carried out in Ref.~\cite{Rinaldi:2013vpa}. Again, the correlations between $x_1$ and $x_2$ are large and the correlations between $x_i$ and $\mathbf{b}$ are small. This framework allows one to gain a clear understanding of the dynamical origin of correlations and to establish which features are model independent. For example, the size of the correlations between $x_i$ and $\mathbf{b}$ is related to relative orbital angular momentum between quarks in the model. Preliminary results obtained in a relativistic light-front scheme indicate that some drawbacks of this calculation can be overcome, such as the so called ``poor support problem''.

\subsection{Charm Cross Sections}

Ref.~\cite{Seymour:2013sya} discusses various issues related to the definition of single and double open charm cross sections. Employing a commonly-used eikonal model, it concluded that $\sigma_\text{eff}$ was too large by a factor of two in the extraction by LHCb. This was recognized in an updated version of LHCb's paper~\cite{Aaij:2012dz} and brings the result for $\sigma_\text{eff}$ from open-charm pairs closer to that from $J/\psi$ plus open charm and jet production. 

The single and double parton scattering production of $c \bar c c \bar c$ and the resulting $D$ meson correlations were studied in Ref.~\cite{vanHameren:2014ava}. The exact $g g \to c \bar c c \bar c$ and $q \bar q \to c \bar c c \bar c$ matrix elements were calculated, leading to a larger cross section at small invariant masses and small rapidity difference between two $c$ quarks than in the high-energy approximation of Ref.~\cite{Schafer:2012tf}. Predictions for the dependence of the cross section on rapidity distance, azimuthal angle and invariant mass of two $D$ meson were compared with recent results from LHCb~\cite{Aaij:2012dz}. The predicted shapes are quite similar but the total cross section is too small. Nevertheless, the calculations in Ref.~\cite{vanHameren:2014ava} clearly confirm the dominance of DPS in the production of events with double charm.

\subsection{Nuclear Collisions}

In nuclear collisions, the effective cross section $\sigma_\text{eff}$ that enters in the pocket formula in \eq{pocket_formula} is modified. In $p$\,-$A$ collisions, the DPS contribution involving different nucleons in the heavy ion $A$ has a nuclear enhanced of $A^{4/3}$, compared to a factor $A$ in when the two collisions involve the same nucleon. Similarly, the contribution involving different nucleons in $A$\,-$A$ collisions is enhanced by a factor $A^{10/3}$, compared to a factor $A^2$ when they involve single nucleons.

Two different LHC final states to which this formalism is applied are: same-sign $W$-boson pair production in $p$\,-$Pb$~\cite{dEnterria:2012qx}, and double-$J/\psi$ production in $Pb$\,-$Pb$~\cite{dEnterria:2013ck}, using NLO predictions with nuclear PDF modifications for the corresponding single-parton scatterings. The first process can help determining the effective $\sigma_\text{eff}$ parameter for $p$\,-$p$ collisions. The second provides interesting insights into the event-by-event enhancements and/or suppressions observed in prompt-$J/\psi$ production in $Pb$\,-$Pb$ collisions at the LHC. Both processes are experimentally measurable and the expected event rates, after acceptance and efficiency losses, for the signal and backgrounds were discussed.

Ref.~\cite{Calucci:2013pza} discussed various aspects of DPS in $p$\,-$Pb$ collisions. First of all, a new class of interferences that arise between different nucleons was pointed out. For $W+2$ jet production these interferences are suppressed, and it was shown that the fraction of events due to DPS in $p$\,-$Pb$ may be a factor 3 or 4 larger than in $p$\,-$p$ collisions at the LHC. This can be extracted from the inclusive transverse spectrum of the lepton produced by the $W$ boson, and depends on correlations between momentum fractions and the typical transverse separation between two partons. The joint study of DPS in high energy $p$\,-$p$ and $p$\,-$A$ collisions, can thus help disentangle the correlation between momentum fractions (approximated as a constant in this study, i.e.~multiplicity correlations) and the transverse separation.

%
%

\section{Progress in the Phenomenology of Double Parton Scattering}

\subsection{Introduction}

The typical signature of double parton scattering (DPS), which is exploited in experimental studies, is the lack of angular and/or momentum correlations between the kinematic properties of the two individual scatterings projected on the transverse plane. The production of four jets is considered to be the most prominent signature for the multiple high-$p_T$ scatterings at hadron colliders, as it may involve two independent scatters in the same collision, each of them producing a jet pair. Contrary to the naive assumption of uncorrelated scatterings, which would basically imply $\sigma_\text{eff} = \sigma_\text{inel}$, the early DPS measurements~\cite{Akesson:1986iv,Abe:1993rv} that inspired the first multiple parton interaction (MPI) models~\cite{Sjostrand:1986ep} seem to favor much smaller values of $\sigma_\text{eff}$.

These early measurements of $\sigma_\text{eff}$ in four-jet events face the major difficulty of vetoing the significant background coming from other sources of jet production, in particular from the QCD bremsstrahlung. That's why the Tevatron strategy to measure directly the hard MPI rates mostly relies on the reconstruction of extra jet pairs in final states with direct photons~\cite{Abe:1997xk,Abazov:2009gc}, which also exploits the better energy resolution performances of photons with respect to jets, resulting in a significant reduction of the combinatoric background. 
The trend of the DPS measurement at the LHC is to favor even cleaner final states with a reduced number of jets or no jets at all. 

In the $4^{th}$ edition of the MPI workshop, ATLAS reported~\cite{Aad:2013bjm} on the production of W bosons in association with two jets in proton-proton collisions at a centre-of-mass energy of $\sqrt{s} = 7$ TeV. Here the DPS component is measured through the $p_T$ balance between the two jets and amounts to a fraction of $0.08\ \pm\ 0.01\ \text{(stat.)} \  \pm \ 0.02 \ \text{(sys.)}$ for jets with $p_T > 20$ GeV and rapidity 
$|y| < 2.8$. This corresponds to a measurement of the effective area parameter for hard double-parton interactions of $\sigma_\text{eff} = 15 \ \pm \ 3\ \text{(stat.)}\ {}^{+5}_{-1}$ (sys.) mb, in agreement with the Tevatron measurements on \mbox{3 jet + $\gamma$} final states. 
In the same edition of the MPI workshop, LHCb reported signals of double \JPsi, double open charm and open charm \JPsi production~\cite{Aaij:2011yc,Aaij:2012dz} in proton-proton collisions at a centre-of-mass energy of $\sqrt{s} = 7$ TeV. 

In this $5^{th}$ edition of the MPI workshop, progress on V + dijet and double \JPsi production in the light of DPS are presented by CMS, that also reports first results on the challenging 4-jet channel. ALICE contributes with a study on open charm and \JPsi multiplicity dependence in pp collisions. This is basically the outline of the next four sub-sections, that includes the summary prepared by the speakers on behalf of their collaboration. 

In general these new experimental results, that show a progress in the comprehension of the relevant systematic uncertainties in particular for what concerns the background modeling, confirm the overall picture of significant DPS rates at the LHC and lay the groundwork for further DPS studies at the LHC, also in the light of understanding unexpected backgrounds in the search for new physics.

In the future, the proposal of much higher center of mass energies (HE-LHC) opens up the possibility to study extreme kinematic regions of QCD that are still not explored. The dynamics of the high energy limit of QCD is driven by the increasingly higher gluon densities at low parton fractional momenta, which should be dominated by multiple parton interactions. With the future high luminosity phase (HL-LHC) it will be possible to achieve very clear evidence of DPS production focusing on the final states with two heavy bosons, in particular on the production of equal sign W boson pairs decaying leptonically, leading to final states with same-sign isolated leptons plus missing energy. 

\subsection{DPS in W + 2-jet final states with CMS at the LHC}

We present a study of DPS based on W + 2-jet events in pp collisions at 7 TeV using CMS detector~\cite{Chatrchyan:2013xxa}.
DPS with a W + 2-jet final state occurs when one hard interaction produces a W boson and another produces a dijet in the same pp collision.
The W + 2-jet process is attractive because the muonic decay of the W provides a clean tag and the large dijet production cross section increases the probability of observing DPS.
Events containing a W + 2-jet final state originating from single parton scattering (SPS) constitute an irreducible background.
Events with W bosons are selected requiring the muon and missing transverse energy information. We also select two additional jets with $\pT >$ 20 GeV/$c$ and $|\eta| <$ 2.0.

\begin{figure}[th]
\centering
\includegraphics[width=2.5in,height=2.5in]{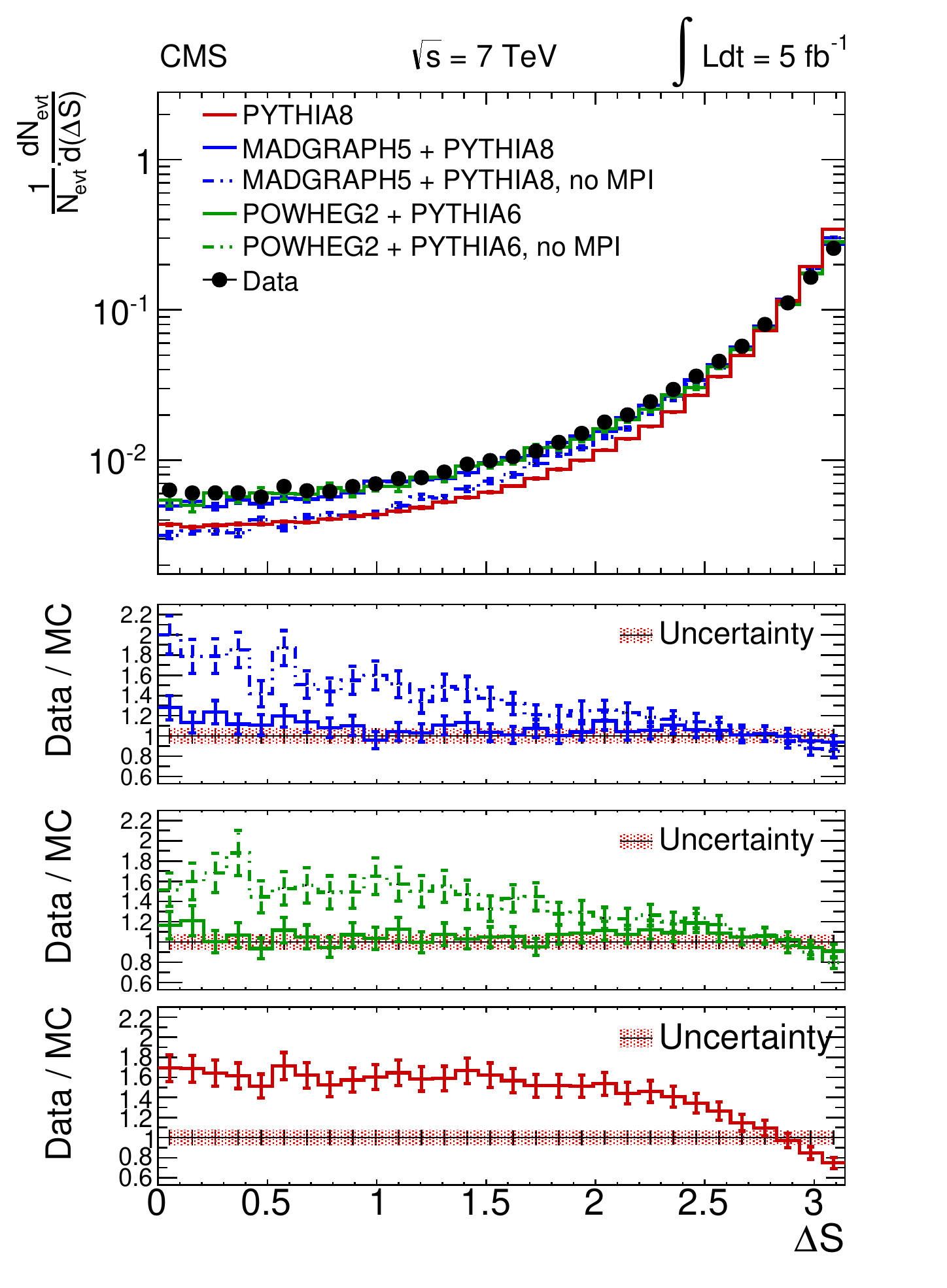}
\includegraphics[width=2.5in,height=2.5in]{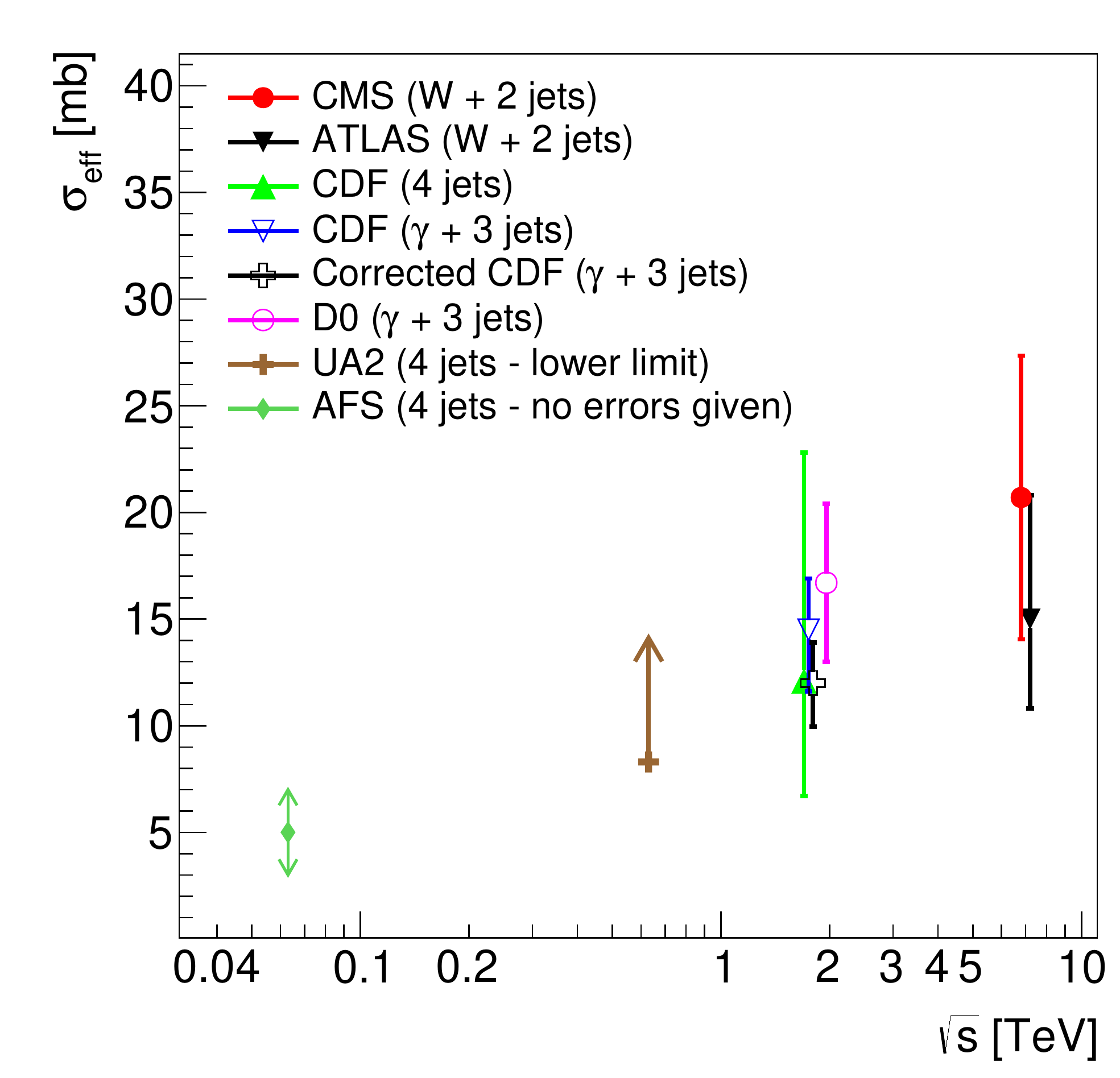}
\caption{  {\it Top}: fully corrected data distributions, normalized to unity, for the DPS-sensitive observable. The second panel in this plot shows the ratio of data over \madgraph5 + \pythia8 with and without MPI, whereas in the third panel the ratio with \powheg2 + \pythia6 is shown.
The ratio of the data and \pythia8 is shown in the fourth panel.
The band represents the total uncertainty of the data. \\
{\it Bottom}: centre-of-mass energy dependence of $\sigma_{\rm eff}$ measured by different experiments using different processes~\cite{Akesson:1986iv,Abe:1993rv,Abe:1997xk,Abazov:2009gc,Aad:2013bjm,Chatrchyan:2013xxa}.
These measurements used different approaches for extraction of the DPS fraction and $\sigma_{\rm eff}$. 
The corrected CDF data point indicates the $\sigma_{\rm eff}$ value corrected for the exclusive event selection~\cite{Bahr:2013gkj}. }
\label{fig:fig}
\end{figure}

We use two uncorrelated observables; the relative $\pT$-balance between two jets ($\Delta^{\rm rel}\pT$) and azimuthal angle between the W-boson and dijet system ($\Delta S$).
The distributions of the DPS-sensitive observables for the selected events are corrected for selection efficiencies and detector effects, and compared with distributions from simulated events. 
Simulations of W + jets events with \madgraph5 + \pythia8(or \pythia6) and NLO predictions of \powheg2 + \pythia6(or \herwig6) describe the data provided MPI are included, as shown in the top panel of Fig.~\ref{fig:fig}. 
\pythia8 standalone fails to describe data mainly in the DPS-sensitive region due to missing higher order contributions.

To extract the DPS fraction, signal and background templates are defined in such a way that signal and background events cover the full phase space.
The fraction of DPS in W + 2-jet events is extracted with a DPS + SPS  template fit to the distribution of the $\Delta S$ and $\Delta ^{\rm rel} p_{T}$ observables.
The obtained value of the DPS fraction is 0.055 $\pm$ 0.002(stat.) $\pm$ 0.014(syst.) and the effective cross section is calculated to be 20.7 $\pm$ 0.8(stat.) $\pm$  6.6(syst.) mb, which is consistent with the Tevatron and ATLAS results, as shown in Fig.~\ref{fig:fig}, and model predictions.

\subsection{Understanding the impact of DPS on V+bb production with CMS at the LHC}

Various measurements have been performed  at the LHC on vector boson production in association with one or more hadrons or jets originating from bottom quarks.
With the increasing luminosities, these studies of V+bb production, that provide a deep test of perturbative QCD, are becoming sensitive to DPS.

The dominant production of Z+bb at the LHC originates from gluon-gluon SPS interactions, a contribution of less than $5\%$, is expected from DPS.
Cross sections of  Z+1b-jets and Z+2b-jets production have been compared with calculations in the four-flavor (4F) and five-flavor (5F) schemes, 
and data turns out to favor \madgraph and aMC@NLO predictions in the 5F scheme~\cite{Chatrchyan:2014dha} including the DPS description.
Differences in kinematics are observed when comparing data and simulation, 
in particular at small angle $\Delta$R$_{{b,b}}$: for this collinear region data favors the 4F prediction from {\sc AlpGen}~\cite{Chatrchyan:2013zja}.

The dominant production of W+bb at the LHC originates from SPS quark-antiquark interactions: here a relatively larger contribution from DPS is expected.
The measurement of the W+2b-jets cross section agrees with theoretical predictions that include a 15\% contribution from DPS~\cite{Chatrchyan:2013uza}.
The measurement of the W+1b-jet production cross section
shows an excess of $\sim1.5\sigma$ with respect to the predictions despite of the fact that the largest expected contribution from DPS are included in the latter~\cite{Aad:2013vka}.

\begin{figure}
\centering
   \includegraphics[width=2.5in,height=2.5in]{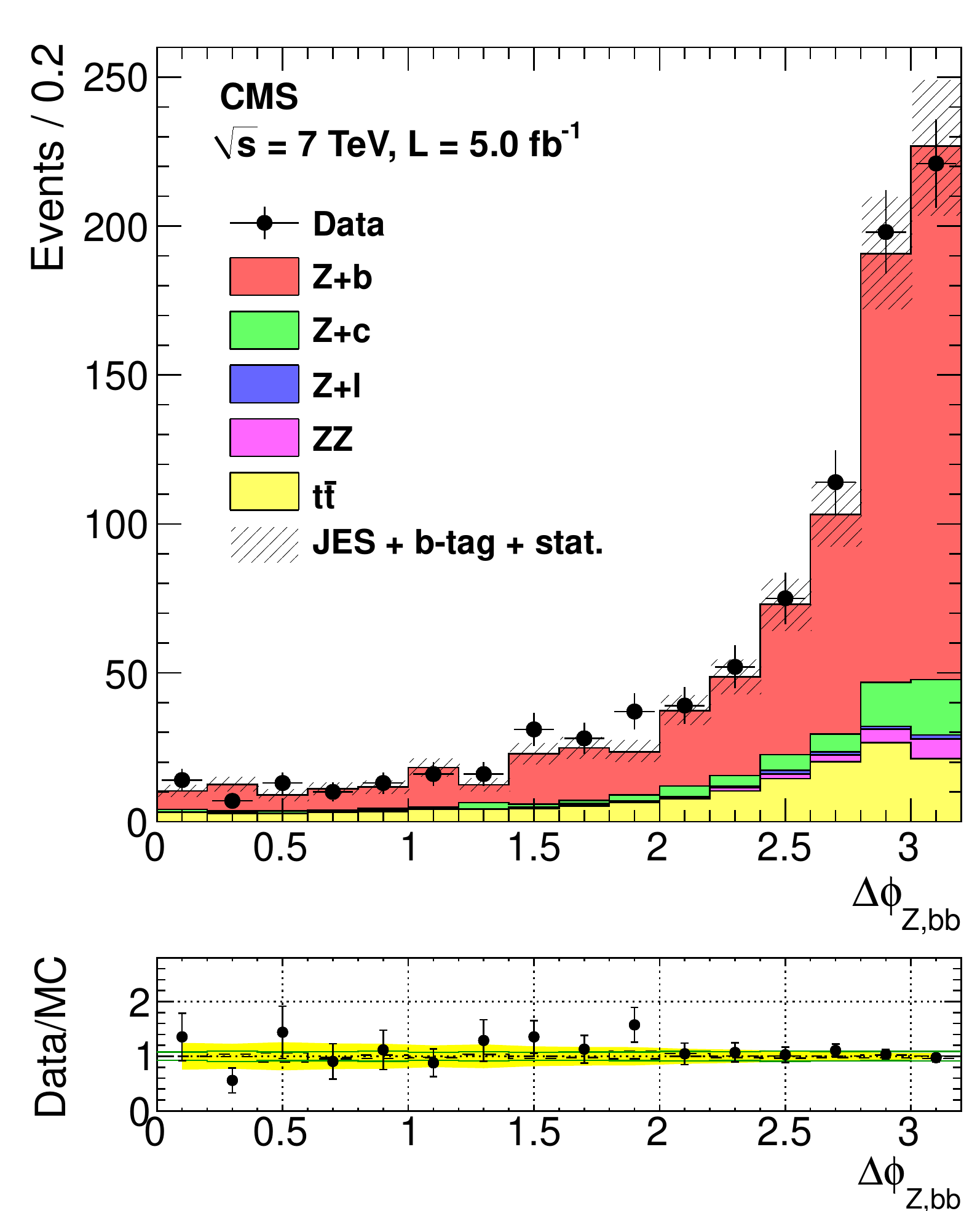}
   \caption{
     Distribution of the azimuthal angle $\Delta \phi_{Z,bb}$ between the Z boson and the dijet system in Z+2b-jets final states~in pp interactions  at $\sqrt{s}=7$~\tev along with the Monte Carlo predictions: Inclusive Z+jets and tt events are simulated with \madgraph 5.1.1.0, using \pythia 6.424 with the Z2 tune for the parton showers, hadronization, and MPIs. The CTEQ6L1 parton distribution functions are used. The ZZ sample is simulated using \pythia. Simulated samples are normalized to the theoretical predictions~\cite{Chatrchyan:2014dha}.
   }
    \label{fig:DeltaPhiZbb}
\end{figure}

Furthermore, DPS-sensitive kinematic properties of the V+bb final states have been studied by CMS using the 7 TeV dataset. 
An example is shown in Fig.~\ref{fig:DeltaPhiZbb}, where the angle $\Delta\Phi_{{Z,bb}}$ shows agreement between data and simulation, within sizable uncertainties. 
The 8 TeV dataset, with reduced statistical uncertainties, will allow for the first measurement of DPS in the V+bb final state.

\subsection{Double \JPsi production with CMS at the LHC}

\begin{figure}[htbp]
  \begin{center}
    \includegraphics[width=2.5in,height=2.5in]{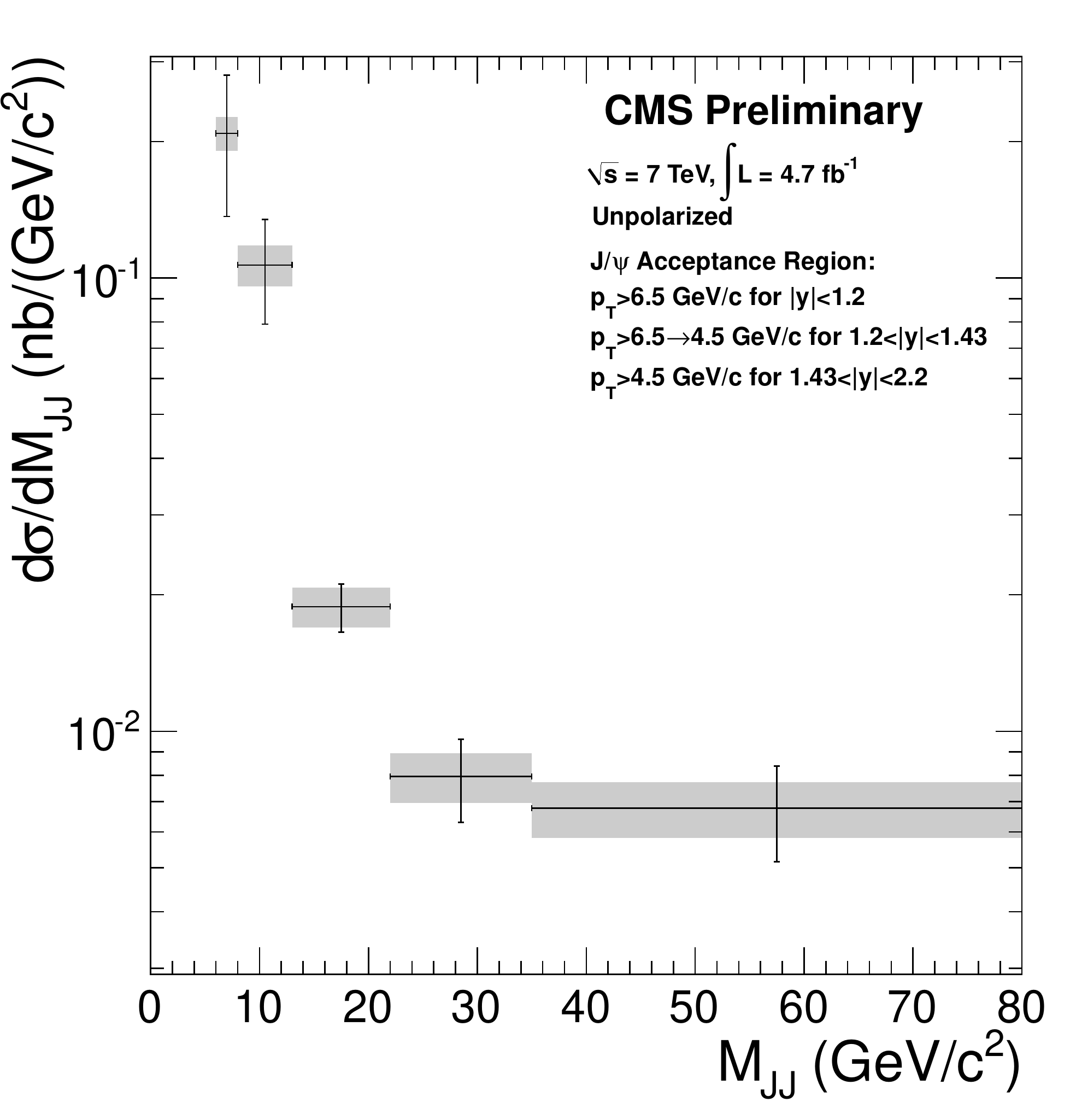}
    \includegraphics[width=2.5in,height=2.5in]{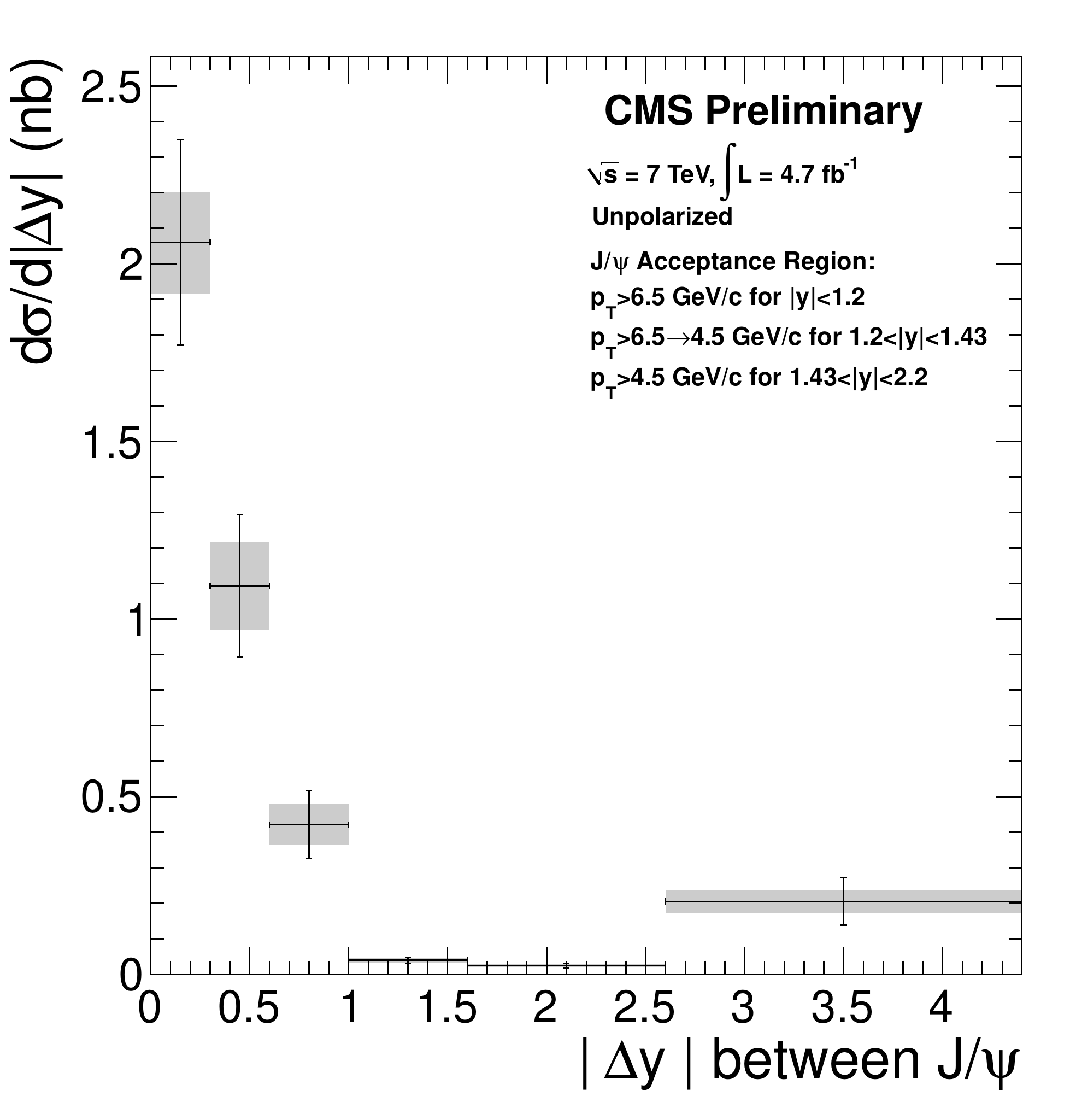}
    \includegraphics[width=2.5in,height=2.5in]{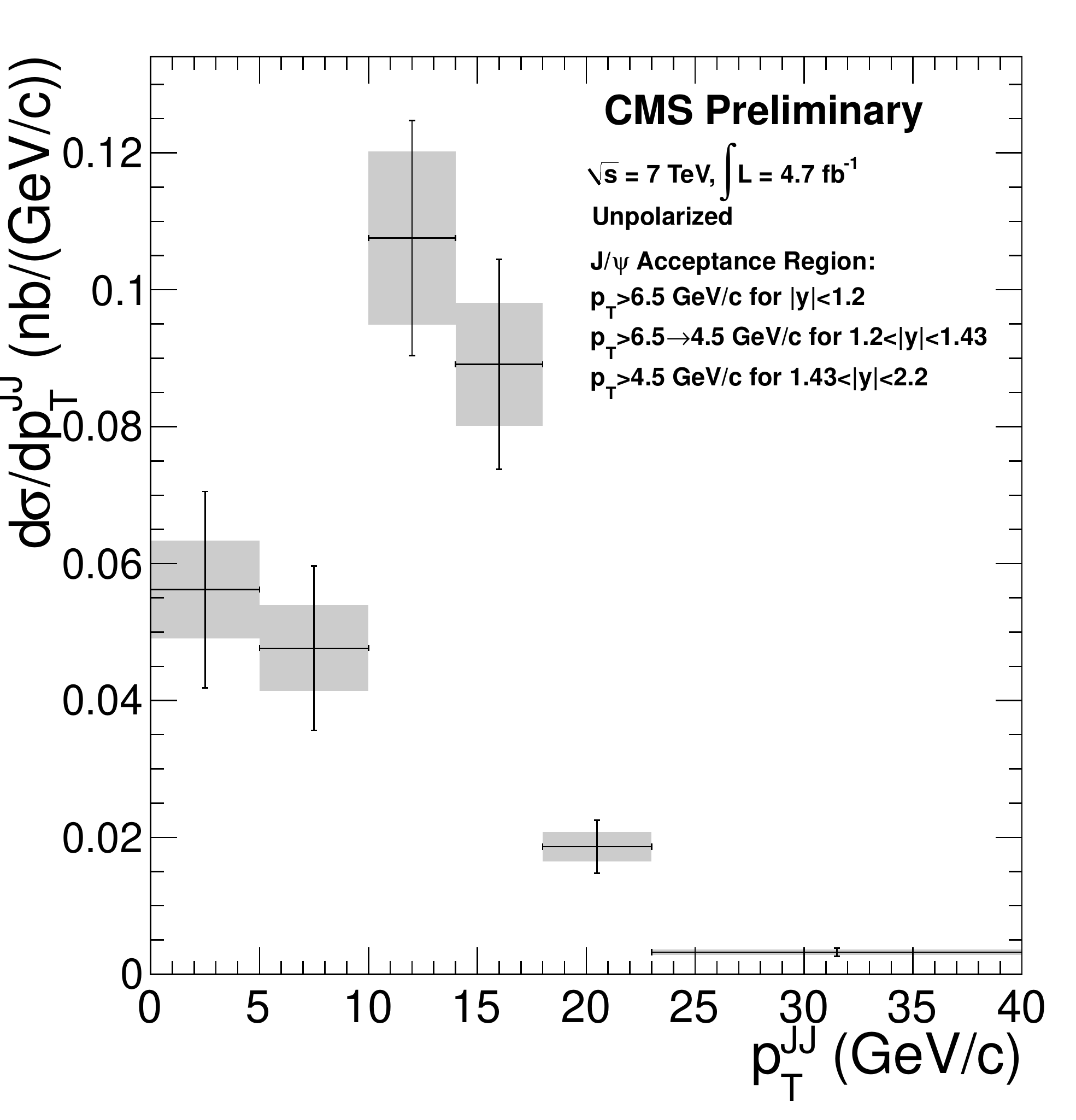}
    \caption{{\it Top:} differential cross section for prompt double \JPsi~production 
as a function of the double \JPsi invariant mass ($M_{\rm JJ}$). {\it Center:} 
absolute rapidity difference between \JPsi~mesons ($|\Delta y|$). {\it Bottom:} double \JPsi~transverse momentum ($\pT^{\rm JJ}$). 
The gray boxes represent statistical errors only, and the error bars represent 
statistical and systematic errors added in quadrature. }
    \label{fig:dsigdx}
  \end{center}
\end{figure}

The goal of this analysis~\cite{Khachatryan:2014iia} is to measure the cross section for prompt double \JPsi~production in pp interactions at $\sqrt{s}=7$~\tev \  
independently from production models. The acceptance corrections rely on the measured \JPsi~kinematics while the efficiency corrections depend on the measured muon kinematics in each event.  
Monte Carlo (MC) samples for different production models with either strongly correlated
\JPsi~mesons (SPS model) or less correlated \JPsi~mesons (DPS model) in 
the event are used to estimate the acceptance region and to validate the 
correction method.
 
At least four reconstructed muons must be identified in an event, forming two pairs of neutrally charged dimuons each with an invariant mass within 250~\mev~of the \JPsi mass.  Signal distributions are extracted from data using an unbinned Maximum Likelihood fit to four event variables: the mass of the \JPsi with higher \pt, the mass of the \JPsi with lower \pt, the proper decay length of the \JPsi with higher \pt, and the separation significance (distance divided by measurement error) between the two \JPsi.  These distributions are parameterized for the signal case using SPS and DPS samples.  
 The dominant background processes are non-prompt \JPsi~mesons (mostly from $B$ meson decays) and combinatorial background from 
a prompt \JPsi~combined with two unassociated muons.  Background components 
and their distributions are extracted from sideband regions in data or 
simulations of \JPsi~mesons from $B$ meson decays. 

A signal yield of $446\pm 23$ events for the production of two prompt
\JPsi mesons originating from a common vertex has been observed 
with the CMS detector in proton-proton collisions at $\sqrt{s}$ = 7~\tev~at 
the LHC from a sample corresponding to an integrated luminosity of 
$4.73 \pm 0.12$~\invfb.   The total cross section of prompt double \JPsi 
production measured within an acceptance region defined by the 
individual \JPsi transverse momentum and rapidity was found to be 
$\sigma = 1.49 \pm 0.07 \pm 0.14$~\nb, where the first uncertainty is 
statistical, the second systematic, and unpolarized production was assumed.  
Differential cross sections were obtained in bins of the double \JPsi 
invariant mass, the absolute rapidity difference between the two \JPsi 
mesons, and the transverse momentum of the double \JPsi system~(Fig. \ref{fig:dsigdx}).  

These measurements probe a higher \JPsi transverse momenta region than previous 
measurements, a region where double \JPsi production via octet \JPsi~states 
and higher order corrections are important.  The differential cross section 
in bins of $|\Delta y|$ is sensitive to DPS contributions to prompt double 
\JPsi production.  The data show evidence for excess at $|\Delta y| > 2.6$, 
a region that current models suggest to be exclusively populated by DPS 
production~\cite{Gaunt:2011rm,Kom:2011bd,Novoselov:2011ff}.
In the double \JPsi invariant mass distribution, no excess above the 
background expectation derived from non-resonant sidebands was observed.

\subsection{DPS in 4-jet final states with CMS at the LHC}

A scenario with four jets at different transverse momentum (\pt) thresholds in the final state has been measured~\cite{Chatrchyan:2013qza} at the CMS detector. In particular, the two hardest jets are required to have \pt greater than 50 GeV and they form the ``hard-jet pair'', while a \pt threshold of 20 GeV is set for the two additional jets, comprising the ``soft-jet pair''. Additional jets with \pt~$>$~20~GeV are vetoed. All the jets are selected in the pseudorapidity region $|\eta|$~$<$~4.7. Single jet \pt and $\eta$ spectra are measured and data are compared to different Monte Carlo predictions, which use a different matrix element in the perturbative QCD framework. All the used predictions include missing higher orders by implementing a parton shower resummation; hadronization and MPI are generated as well. The \pt spectrum of the hard jets is well reproduced by all the compared models at high \pt, while at low \pt, where a contribution coming from MPI is expected, the predictions start to exhibit greater differences. The \pt distributions of the soft jets are only reproduced by some models. Absolute cross sections and normalized distributions of correlation observables between the jet pairs are also measured. The comparison of the normalized differential cross sections as a function of the correlation observables shows that the present calculations agree only in some regions of the phase space and the contributions from SPS can be improved by higher order calculations. In addition, the predictions including MPI need to be validated with underlying event measurements before a direct extraction of the DPS contribution can be performed. However, the measurements may be taken as an indication for the need of DPS in the investigated models.

\subsection{Open charm and \JPsi multiplicity dependence in pp collisions at
$\sqrt{s}$ = 7 TeV with ALICE at the LHC}

ALICE, that already reported \cite{Abelev:2012rz} the production of \JPsi mesons as a function of the charged particle multiplicity in proton-proton collisions at $\sqrt{s} =$ 7.0 TeV, is currently extending these correlation studies to the open charm yields at central 
rapidity ($|y| < 0.9$) and to the \JPsi mesons 
at both central and forward (2.5 $< y <$ 4) rapidity, in pp 
collisions at $\sqrt{s} = 7$ TeV. 

Concerning the \JPsi analysis, at central rapidity 
the fraction of \JPsi 
coming from the decay of beauty hadrons, $f_{\rm B}$, is also measured,
providing an estimation of the multiplicity dependence of beauty hadron 
yields. The comparison  with the D meson yields suggests, within the present 
uncertainties, a similar behavior of charm and beauty hadron 
production as a function of the event multiplicity. 

A Monte Carlo study using the \pythia 8 simulations was 
performed: several contributions for charm and beauty 
particle production were disentangled and their behavior 
was studied as a function of the charged particle multiplicity.
In particular the rate of flavor creation via gluon fusion ($\rm q\bar{q}$ 
annihilation, with hard interactions involving sea quarks) 
show a flat behavior as a function of the event multiplicity. 
Contributions from gluon splitting from
initial and final state radiation (ISR/FSR) and from hard processes 
in multi-parton interactions, instead, show a linear increase as a function of 
multiplicity, similar to that observed in data. This gives a hint
of MPI as a possible source for the observed linear increase of
the yields, but no conclusion is possible considering 
the present uncertainties. Further Monte Carlo studies are actually ongoing and the final results are expected to be published in an ALICE paper in preparation along with the results of the D mesons and non-prompt \JPsi multiplicity dependence measurements.

\bibliography{mpi_2013_dps}

\end{document}